# Identifying and Managing Technical Debt in Database Normalization Using Machine Learning and Trade-off Analysis

Mashel Albarak
University of Birmingham UK
King Saud University KSA
mxa657@cs.bham.ac.uk

Muna Alrazgan
King Saud University
KSA
malrazgan@ksu.edu.sa

Rami Bahsoon
University of Birmingham
UK
r.bahsoon@cs.bham.ac.uk

## ABSTRACT

Technical debt is a metaphor that describes the long-term effects of shortcuts taken in software development activities to achieve near-term goals. In this study, we explore a new context of technical debt that relates to database normalization design decisions. We posit that ill-normalized databases can have long-term ramifications on data quality and maintainability costs over time, just like debts accumulate interest. Conversely, conventional database approaches would suggest normalizing weakly normalized tables; this can be a costly process in terms of effort and expertise it requires for large software systems. As studies have shown that the fourth normal form is often regarded as the "ideal" form in database design, we claim that database normalization debts are likely to be incurred for tables below this form. We refer to normalization debt item as any table in the database below the fourth normal form.

We propose a framework for identifying normalization debt. Our framework makes use of association rule mining to discover functional dependencies between attributes in a table, which will help determine the current normal form of that table and identify debt tables. To manage such debts, we propose a trade-off analysis method to prioritize tables that are candidate for normalization. The trade-off is between the rework cost and the debt effect on the data quality and maintainability as the metaphoric interest. To evaluate our method, we use a case study from Microsoft, AdventureWorks. The results show that our method can reduce the cost and effort of normalization, while improving the database design.

## CCS CONCEPTS

• **Software and its engineering** → *Designing software;*

• **Information systems** → **Relational Database Model;**

## KEYWORDS

Technical debt; Database design; Normalization; Data mining

## 1 INTRODUCTION

The metaphor of technical debt is used to describe and quantify issues arising from system evolution and maintenance actions taken to modify the software, while compromising long-term qualities for immediate savings in cost and effort [9]. The analysis of technical debt involves trade-offs between short term goals (e.g. fast system release; savings in Person Months) and applying optimal design and development practices [21]. There has been an increasing volume of research in the area of technical debt. The majority of the attempts have focused on code and architectural level debts [21]. However, technical debt linked to database normalization, has not been explored, which is the goal of this study. In our previous work [4], we defined database design debt as:

> " The immature or suboptimal database design decisions that lag behind the optimal/desirable ones, that manifest themselves into future structural or behavioral problems, making changes inevitable and more expensive to carry out".

In [4], we developed a taxonomy that classified different types of debts which relate to the conceptual, logical and physical design of the database.

In this study, we explore a specific type of database design debt that fundamentally relates to normalization theory. Database normalization is one of the main principles for relational database design invented by the Turing Award winner, Ted Codd [8]. The concept of normalization was developed to organize data in tables following specific rules to reduce data redundancy, and consequently, improve data consistency by decreasing anomalies. Benefits of normalization go beyond data quality and can have ramifications on improving maintainability, scalability and performance [19], [23], [12]. Conventional approaches often suggest higher level of normalization to achieve structural and behavioral benefits in the design. Despite the claimed benefits from this exercise, some database designers neglect the normalization process due to the time and expertise it requires, and instead, turn to other procedures to achieve quick benefits. These procedures resemble in writing extra code fixes in the software to maintain data consistency, or create extra indexes to improve performance. However, these quick and ill-practices can have negative impacts on the system, accumulating to call for costly future rework, replacements and/or phasing out [15], [4].

### 1.1 Novel Contributions and Research Questions

The novel contribution of this paper is a method that can help software engineers and database designers to re-think normalization exercise and their design decisions not only from their technical merits, but also from their connection to debt. The contribution will help engineers to identify the likely debt in database normalization and consequently work on managing these debt(s). Coining normalization with debt hopes to provide a systematic procedure that can eliminate unnecessary normalization efforts, justify essential ones, and/or prioritize investments in normalization, when resources and budget are limited.

We aim to answer the following questions:
- What is database normalization debt?
- How to identify database normalization debt?
- How to manage database normalization debt?
- To what extent normalization debt can provide designers with extra insights to the exercise? In particular, how Normalization Debt can justify the normalization design decisions from the cost and debt perspectives?

A prerequisite for managing technical debt is to make the debt explicit, that is, to identify it. Some technical debt types can be identified through static program analysis tools that detect anomalies in the source code [34]. However, such tools are not available to detect the flaws in the database design. Our method uses a data mining technique, association rule mining in particular, to identify database normalization debt items. Since fifth normal is often regarded as a theoretical form [15], we refer to normalization debt item as any table below the fourth normal form. To identify the debt items, the technique discovers the relationship between the attributes in each table, referred to as "functional dependencies" between the attributes, which will reveal tables below fourth normal form.

After identifying the debt items, the next step toward managing such debts is to quantify the principal. The method to estimate the principal of that debt considers the rework cost needed to resolve the debt and normalize the table to the 4th normal form. Normalization involves decomposing a table to several tables to satisfy the normal form criteria. This alteration to the database design can be costly as it will also involve a lot of refactoring tasks to be done on the database schema, and the applications making use of the database as well. Models have been developed to estimate the cost of software refactoring, such as [27]. However, up to our knowledge, the cost of refactoring databases, specifically normalizing tables, and the consequences of such modifications on the schema and application programs, has not yet been explored, which is one of this study aims.

Finally, the decision to pay off the debt requires more information on how the debt affects the system. In this research, we use two quality attributes, maintainability and data quality, as metrics to quantify the debt influence, which resembles the interest rate one currently pays on the debt. These metrics are later used in a trade-off analysis between the quality impact (i.e. debt interest) and the rework cost of normalization (i.e. debt principal) to prioritize tables need to be normalized.

The techniques are evaluated using a case study from Microsoft, AdventureWorks database [2] and StoreFront web application built on top of the database [28]. The database has a total of 64 tables, each populated with large amount of data to support e-commerce transactions of a hypothetical cycles retail company. The size of the database and the amount of data approximate the scale of complex real life systems.

Our method has the promise to replace ad hoc and uninformed practices for designing databases, where debt and its principal can motivate better design, optimize for resources and eliminate unnecessary normalization.

## 2 BACKGROUND AND MOTIVATION

In this section, we will discuss the key concept of our work. The normalization process was first introduced in 1970 [8], as a process of organizing data in tables following specific rules to satisfy each normal form criteria. Fig 1 illustrates the normal forms hierarchy. The higher the level of normal forms, the better design will be [15], as higher levels will reduce more redundant data. The main condition of moving from one normal form to the higher level is based on the dependency relationship, which is a constraint between two sets of attributes in a table. The main goal of normalization is to reduce data duplication, which is accomplished by decomposing a table into several tables after examining the dependencies between attributes. This principle has many advantages [12], [15] such as: improve data consistency as it reduces data redundancy; facilitate maintenance by decreasing tables complexity and even improve performance as proven in the literature [23], [29], [12]. Therefore, weakly normalized tables in the database will negatively affect the quality of the system in various dimensions. In this study, we viewed these effects and complications as symptoms of a debt carried by not normalizing the database tables. These complications will accumulate overtime with system growth and increasing stakeholders' requirements. It is evident that the short-term gains from not normalizing the table can have long-term consequences that can call for inevitable costly maintenance, fixes and/or replacement of the database. Therefore, we motivate rethinking database design and engineering from debt perspective that is attributed to normalizing issues in the database. Rethinking normalization from the technical debt angle can lead to software that has potential to better evolve in the face changes.

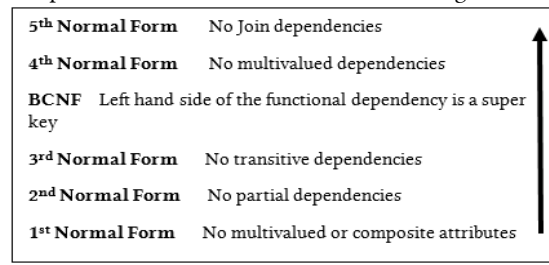

**Figure 1 Normal forms hierarchy**

Automated database normalization captured the attention of many researchers [14], [13]. Their approaches focused on developing algorithms to produce third normal form or BCNF tables automatically. However, their studies looked into the schema in isolation from applications using the database, meaning they did not consider the rework cost of normalization taking into account all the refactoring, configuration, and data migration tasks. Since this process can be costly, our work takes into consideration the added value and also the rework cost of normalization. Our aim is to balance this cost with the value of normalization to improve the database quality. Database schema evolution is well studied in literature [10], [7]. However, researchers haven't looked at the evolution for the purpose of normalizing the database to create value and minimize debts. Evolution should be done to create a value. By creating a value, it is ensured that the system will sustain and be much more usable and maintainable. On the other hand, if value is not created while carrying too much debt, it is likely that we have to think of the evolution procedure. Our study aims to explore the impact and the cost of normalizing a database on the system as whole rather than focusing on the evolution method alone.

## 3 NORMALIZATION DEBT DEFINITION

There is no general ageement on the definition of technical debt [20]. Existing definitions have steered by various research perspectives and their pending problems for technical debt identification, tracking, and management etc. However, The common ground of these definitions is that the debt can be attributed to poor and suboptimal engineering decisions that may carry immediate benefits, but are not well geared for long-term benefits. Database normalization had been proven to improve the quality in various dimensions [12]. Consequently, tables below the $4^{th}$ normal form are subjected to potential debts as they potentially lag behind the optimal, where debt items may cause problems of data inconsistency and might decrease overall quality as the database grows [12], [23]. To address this phenomenon of normalization and technical debt, we have chosen $4^{th}$ normal form as our target normal form. Although most practical database tables are in $3^{rd}$ (rarely they reach BCNF) normal form [15], $4^{th}$ normal form is considered to be the optimal target since more duplicated data is removed. Additionally, 4th normal form is based on multi-valued dependency which is common to address in real world [32]. Moreover, higher normal form of the $5^{th}$ level is based on join dependencies that rarely arise in practice, which makes it more of a theoretical value [12], [15].

Following similar ethos of database design debt definition [4], we view that database normalization debt is likely to materlise for any table in the database that is below the $4^{th}$ normal form.

## 4. NORMALIZATION DEBT IDENTIFICATION

Given the previous definition of normalization debt, our objective is to identify debt tables that are below the $4^{th}$ normal form by determining the current normal form of each table in the database. Determining the normal form can easily be done through knowledge of the functional dependencies that holds in a table. A functional dependency is a constraint that determines the relationship between two sets of attributes in the same table [12]. These dependencies are elicited from a good knowledge of the real world and the enterprise domain, which the database will serve. Experienced database developers and the availability of complete documentation can also be a source for extracting dependencies. Practically, none of these might be available, which makes dependency analysis a tough task. We thus propose a framework to determine the current normal for each table by mining the data stored in each table. The framework will make use of a data mining technique called association rules [26] to analyze the data and test candidate dependencies for each table.

Association rule mining is a machine learning method that discovers interesting relationships between variables in a big set of data [26]. The objective of this method is to search for strong rules using some measures of interestingness. The discovery of such rules provides insights to make better business decisions. A famous example of association rule technique is the market basket data analysis. The process involves analyzing customers buying habits by searching for association between different items that the customers purchased. For example, customers who are buying ice cream also tend to buy candy and water at the same time. This can be represented in association rule as follows:

Ice cream → candy, water   [Support = 0.05 confidence = 1]

The support of 0.05 for the rule means that 5% of all the transactions under analysis showed that ice cream, candy, and water are bought together. The confidence of 1 means that 100% of the customers who bought ice cream also bought candy and water. The confidence metric is automatically calculated by the algorithm as the following:

Confidence(X→Y) = Support(X∪Y)/Support(X).

We have adopted the same approach in our aim of identifying normalization debts. Association rules will facilitate the process of revealing functional dependencies that violate and those that meet normal forms criteria, For example, table `Employee` stores information about employees in a certain company has the following attributes`(Id,job_title, Salary, age,…)`. After mining the data stored in this table, the following rule would be one of the rules extracted:

Job_title:Admin → Salary:10K   [confidence =1]

Since the confidence of this rule is equal to 1, the rule can be interpreted as: all Admins have same salary which is 10K. Similar rules for all job titles would also be found after the mining process. These rules indicate the relationship between Job_title and Salary attributes in the table, and depending on the keys of the table, we can determine the normal form of this table. If the confidence of any rule found to be less than 1, it would indicate that the functional dependency: JobTitle→Salary, does not hold in the table and we have to test another functional dependency between two different sets of attributes.

Our framework is composed of several steps to determine the normal form for each table. Fig 2 displays all the steps involved in identifying the debt. At first, since $1^{st}$ normal form is the foundation of the relational theory, it should be confirmed before using the framework. Tables with too many columns are an indication of a lack of cohesion and storing data for multiple entities [5]. For example, an Employee table that has four columns to store four different addresses. It is likely that this table is not normalized even to the $1^{st}$ normal form. In the framework, we used one of the association rule techniques, which is Frequent Pattern or simple FP--growth algorithm to discover the associations among attributes and test dependencies related to each normal form. FP--growth adopts a divide-and-conquer strategy [17]. It allows frequent pattern discovery efficiently by first, dividing the data and building a compact data structure called the FP- tree. Then, extract "rules" directly from the tree. In Fig 2, the blue rectangles depict where the mining is performed and the red circles determines when to stop the process and identify the normal form of the debt table. The green circles are also ending points and determine the table is in 4th normal form. Prior the mining step, potential functional dependencies violating the criteria of the normal form are identified based on any data available from the data dictionary or a simple knowledge of the domain that the database is serving.  After that, the mining process is performed to test those dependencies if they apply in the table or not. The confidence of the rules is used to determine if a functional dependency holds in this table. A confidence less than 1 would mean that this dependency does not hold in the table. Otherwise, a confidence of 1 means that this dependency holds in the table which violates the normal form. Resulting from the association rules, if a violation is found, the process should stop and the table would be in the weaker normal form. If no violations were found after testing all potential dependencies, the process should continue to examine the higher normal form. Detailed explanation of the framework will be illustrated via a case study in the next section.

## 4.1 Case Study

We considered the AdventureWorks database, designed by Microsoft [2]. This database supports standard e-commerce transactions for a fictitious bicycle manufacturer. The database has a total of 64 tables, each filled with a thousands and hundreds of fictitious data. The data dictionary is available with a fair description of the tables and the attributes, which facilitates some of the steps such as, identify the candidate keys and some potential dependencies. To better understand our framework, the following example demonstrates how the steps are performed to identify the table current normal form.

### 4.1.1 Example: table `Address`

We will follow the framework steps on table named `Address` that has the following attributes (AddressID, Addressline1, Addressline2, City, State ID, PostalCode).

**Step 1**: Identify all candidate keys of the table: a candidate key is an attribute or composite set of attributes that uniquely identifies a tuple in a table. Candidate keys must be unique, not null and irreducible [12]. This step will facilitate the process of discovering candidate dependencies related to each normal form in the subsequent steps. Several researchers have proposed algorithms to discover candidate keys for large databases [1]. The simplest approach is by issuing a query to retrieve unique values of an attribute (or composite attributes), and then compare the number of the retrieved unique values to the total number of rows in the table. If they are equal then this attribute (or composite attributes) is a candidate key. This process is iteratively tested among all the attributes. In the `Address` table, aside from the primary key, one single candidate key is found: AddressLine1. Since no composite keys were found, the table is considered to be in the 2nd normal form and we will continue to the next step.

**Step 2a:** test dependencies between non-key attributes to determine 3rd normal form. 3rd normal form is based on transitive dependencies between non-key attributes. Thus, we need to first identify candidate dependencies between non-key (or a set of non-key) attributes. Potential transitive dependencies in the `Address` table would be between (City, PastalCode, StateID). After running the FP-Growth algorithm on those attributes, we looked for any rule with a confidence less than 1 in the table. For example, we found the following rule:

PostalCode = V63P7 → City = Richmond [Confidence=0.105]
This implies that there is a different city for the same PostalCode, which means that the functional dependency Postalcode → City does not hold in the table. Similarly, other transitive dependencies were not found in the table, which implies that it is in 3rd normal form, and even in BCNF normal form sine it does not have any composite keys.

**Step 2a-1**: Test multivalued dependency to determine 4th normal form: 4th normal form is based on multivalued attributes dependencies. Testing this kind of dependency is done manually through inspecting the data and testing embedded dependencies as well [12] as this is not possible to achieve via association rules . We found that the following multi-valued dependencies hold in

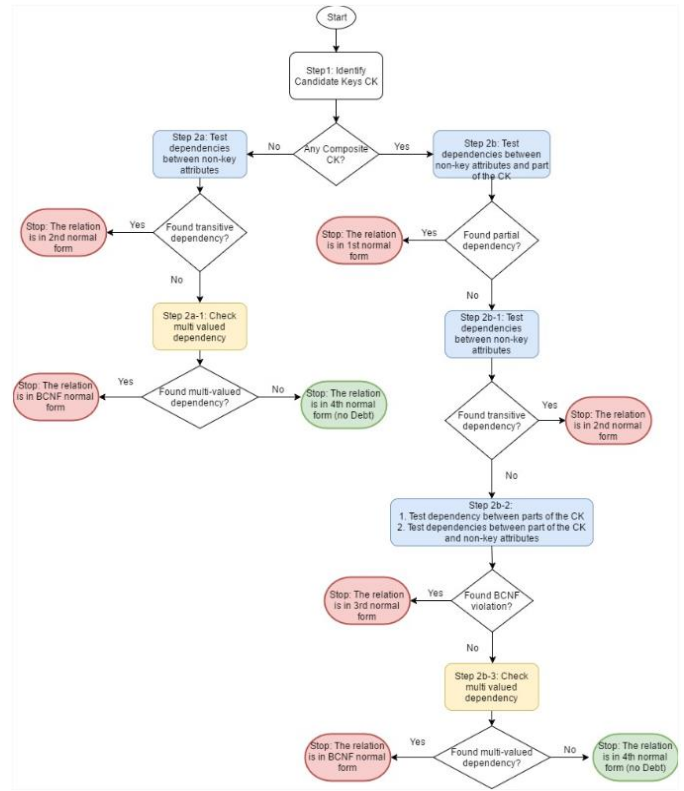

**Figure 2:** **Normalization debt identification framework**

the table: City →→ StateId ; City →→PostalCode
Such rules violate 4th normal form criteria. Therefore the table is in BCNF and it considered as a debt item.

We have applied the framework on all the 64 tables, and were able to identify 23 debt items of tables below 4th normal form. Due to space limitation, we will present eight of these tables. Table 1 displays the debt items with their normal form.

**Table 1:** AdventureWorks Debt Tables

| Table Name | Normal form |
|---|---|
| Address | BCNF |
| Employee | 1st Normal Form |
| EmployeePayHistory | 1st Normal Form |
| Product | Un-normalized |
| ProductProductPhoto | 1st Normal Form |
| SalesTaxRate | BCNF |
| Vendor | 2nd Normal Form |
| WorkOrder | 2nd Normal Form |

Referring to the previous table, technical debt can vary across the tables. Our intuition is that each table can have varying effect on the system regardless of its normal form level; That is to say, there can be a higher normal form table with more negative impact on the system and carrying more debts than the weaker normal form ones. We will confirm our intuition in the next section.

## 5 MANAGING DATABASE NORMALIZATION DEBT

After identifying debt items and making them explicit to the developers, additional information is required to manage those debts. In general, managing technical debt is choosing between paying off the debt, in our case normalize the table to 4th normal form, or keep the debt unresolved with regular monitoring. As mentioned previously, normalization level cannot be the only criteria to make that decision. In the case study for example, normalizing all of the debt tables is very time and money consuming and refactoring the whole system might be cheaper and more efficient. Therefore, a more proper approach is to prioritize debt items the need to be normalized. A logical strategy would be to normalize tables that are inexpensive to split and have the biggest negative impact on the database quality. Accordingly, tables that are expensive to fix and only promise small, or no, gain in database quality is deferred. Therefore, an approach for prioritizing normalization debt should incorporate both dimensions:

- Debt Principal: the cost of normalization and
- Quality impact of debt items.

### 5.1 Estimating Normalization Debt Principal

The principal of a technical debt is the cost to be paid to eliminate that debt [11]. That is, the effort required to resolve the debt by changing or refactoring the software. Principal estimation is one of the bases of technical debt management [16],[21].The estimation process is essential prior making the decision to pay the debt and refactor the system. This principal must conform to the company's budget and time constraints to avoid potential loss.

Principal of database normalization debt is the effort cost to normalize the table to the 4th normal form. Normalizing a table involves decomposing a given table to two or more tables to achieve normal form criteria. Splitting a single table in the database requires several refactoring tasks[5]:

- Modifications to the database schema
  - The new table/s after decomposition must be created.
  - To ensure that the data in the new tables will be synchronized with one another, triggers should be created. These triggers will be dropped at some point after committing all the changes.
  - Extract and update all the SQL statements embedded in the application code, functions, procedures and database views.
- Migrate data to new table/s
- Modifications to the accessed application/s:
  - Introduce the new table meta-data .
  - Refactor the presentation layer: the user interface should reflect the fine-grained data changes.

Some of these tasks can be automated depending on the technologies used in the project. However, regardless of the size of the system, refactoring the database for normalization can be very complex and time consuming. Therefore, we propose a model to estimate normalization debt principal for a single debt table as:

Principal= $\sum_{i=1}^{n}$ (average hours needed to complete the *i*th refactoring task × labor cost in $ per hour).

Where n is the total number of refactoring tasks needed to normalize the table to the 4th normal form, migrate the data, update the database schema and applications using the database. The number of tasks and time required to perform the refactoring tasks can be estimated from historical effort data and domain experts.

To illustrate the method, we will consider table `Product`. This table is un- normalized, thus considered as a debt table. After examining the table, it is estimated that it needs to be decomposed to five tables to achieve 4th normal form criteria. Manual inspection of the database and the application was done, and the total number of refactoring tasks is estimated to be 50 tasks. For simplicity, we will assume that each task will take an approximate five to ten minutes to perform, as anticipated for software refactoring tasks in [31]. We realize that practically, in database refactoring some of the tasks will take longer time, however, this is only an assumption to clarify the method. The mean hourly wage for software developers/systems software is estimated to be $51.38 based on the US Bureau of labor statistics report [36], leading to a rough estimate of 6.42$ per refactoring task. Therefore, to estimate the cost of normalizing table `Product` to the 4th normal form, multiply the total number of the refactoring tasks by 6.42 $, which equals to 321 $. Similarly, total cost for paying the debt and normalizing some of the tables to 4th normal form is calculated and illustrated in Table 2.

**Table 2:** Principal of Debt Tables

| Table name | Total cost in $ |
|---|---|
| Employee | 102.72 |
| EmployeePayHistory | 57.78 |
| ProductProductPhoto | 70.62 |
| WorkOrder | 51.36 |
| Vendor | 64.2 |
| SalesTaxRate | 51.36 |
| Address | 77.04 |

## 5.2 Quantifying Quality impact of Normalization Debt (Debt Interest)

One crucial factor for managing debt is to consider the interest on that debt. Unlike finance, interest on technical debt has been acknowledged to be difficult to elicit and measure for software engineering problems [16]. The interest can span various dimensions; it can be observed on technical, economic, and/or sociotechnical dimensions etc. For this work, we liken the interest to the quality impact of the debt tables on the system. The decision to pay off normalization debt considers the impact of debt tables on the quality (i.e. interest). Quality attributes are among the determinants for stakeholders' satisfaction and the extent to which the system meets its specification [24]. For this study, we measure the interest by considering the implications of the debt tables on two quality attributes described below. Nevertheless, the approach can consider other attributes:

- Data quality: represented by risk of data inconsistency.
- Maintainability: represented by tables' complexity and size.

### 5.2.1 Risk of data inconsistency

According to the International Organization for Standardization (ISO/IEC) [18], risk of data inconsistency is the risk of having inconsistency due to data redundancy. This risk is proportional to the number of duplicate values as each update process should be performed to all occurrences of the same value to avoid inconsistency. The risk of data inconsistency is decreased if the table is further normalized to higher normal forms and more redundant data is eliminated [18]. This measure may provide more insight about the level of the debt for each table that shouldn't rely on the level of normalization alone. Meaning, some tables might be in a low normal form but also have low risk of data inconsistency because it stores low amount of redundant data. On the other hand, higher normalized table might have greater risk of data inconsistency due to the huge amount of duplicate data it stores. Data inconsistency risk for a single table is measured according to the following formula [18]:

$X = A/B$

With $\binom{n}{k}$ sets of k attributes for a table with n attributes (k=1,..n),

$A = \Sigma_k \Sigma_i D_i$

$D_i$ = number of duplicate values found in set i of k attributes.

B= number of rows × number of columns.

Rows refer to the data stored in the table and columns refer to the attributes of that table. For X, lower is better. In the case study, we used this metric to determine the risk of data inconsistency for the tables. The following Table 3 displays the normal form for some of the tables in the database and the risk of data inconsistency for those tables. As seen from the table, the risk of data inconsistency could be higher in highly normalized tables than weakly normalized table. For example, table `Address` is in BCNF and the risk of data inconsistency is higher than table `ProductProductPhoto` which is in 1st normal form. This emphasizes our argument that normal form alone is not sufficient to make the proper decision and normalize the table

**Table 3:** Risks of Data Inconsistency of Debt Tables

| Table Name | Normal form NF | Risk data inconsistency | Number of rows |
|---|---|---|---|
| Product | Un-normalized | 17 | 504 |
| Employee | 1st NF | 8.213 | 290 |
| EmployeePayHistory | 1st NF | 1.034 | 316 |
| ProductPhoto | 1st NF | 0.978 | 504 |
| WorkOrder | 2nd NF | 2.910 | 72591 |
| Vendor | 2nd NF | 0.987 | 104 |
| SalesTaxRate | BCNF | 0.8 | 29 |
| Address | BCNF | 1.159 | 19614 |

### 5.2.2 Tables' Complexity and size

Tables' complexity and size will provide more insight on normalization debt effect on the database quality. In this study we compute the complexity for some of the tables using the Database Complexity method (DC method) [25]. The DC method calculates the weight for each table in a database as a sum of: number of attributes, number of indexes, number of foreign keys, and number of keys. The complexity of the database will then be the total sum of the table weights. This complexity can be decreased through normalizing tables as decomposition will reduce the number of attributes and the amount of duplicate data stored in each table. The following Table 4 displays the weight of some tables in the AdventureWorks case study. It will also display the size of those tables in megabytes. By size we mean the space allocated in the disk by the data stored in each table.

**Table 4:** Debt Tables' complexity and size

| Table Name | Normal form NF | Weight (complexity) | Size in MB |
|---|---|---|---|
| Product | Un-normalized | 28 | 0.102 |
| Employee | 1st NF | 21 | 0.055 |
| EmployeePayHistory | 1st NF | 7 | 0.016 |
| ProductProductPhoto | 1st NF | 6 | 0.016 |
| WorkOrder | 2nd NF | 10 | 4.125 |
| Vendor | 2nd NF | 10 | 0.016 |
| SalesTaxRate | BCNF | 9 | 0.008 |
| Address | BCNF | 11 | 2.719 |

From the previous table, it is clear that tables' complexity is independent from its normal form. For example, table `Address` in BCNF normal form weights more than `EmployeePayHistory` table which is in a weaker normal form. Therefore, this emphasizes that quality measure provides more insight on the level of the debt and its effect on the system, which will lead to a better decision about database normalization.

### 5.3 Prioritizing Normalization Debt

In previous sections, we identified the debt items, estimated the debts' principals and measured the impact of each debt on the quality. The next step is to incorporate these information to prioritize tables that need to be normalized. The proposed method use trade-off analysis between the cost of normalizing to the ideal form and the current quality impact of the debt items. This is achieved by ranking both cost and the quality impact of the tables. The following Table 5 demonstrates ranking of some AdventureWorks tables based on the cost of normalization from cheapest to most expensive.

**Table 5:** Normalization Cost Ranking

| Table name | Rank |
|---|---|
| WorkOrder | 1 |
| Sales Tax Rate | 1 |
| Employee Pay History | 2 |
| Vendor | 3 |
| ProductProduct Photo | 4 |
| Address | 5 |
| Employee | 6 |
| Product | 7 |

Ranking of the quality impact for the same tables from lowest negative impact to highest is shown in Table 6. First, ranking of tables' impact on the three quality metrics is determined based on the information detailed in section 5.2. After ranking all three metrics, the total rank will be calculated by adding the three ranks for each table and ranking this result.

**Table 6:** Debts' Quality Impact Rank

| Table name | RDI Rank* | TC Rank* | TZ Rank* | Sum of ranks | Overall Rank |
|---|---|---|---|---|---|
| WorkOrder | 6 | 4 | 6 | 16 | 5 |
| SalesTax Rate | 1 | 3 | 1 | 5 | 1 |
| EmployeePay history | 4 | 2 | 2 | 8 | 2 |
| Vendor | 3 | 4 | 2 | 9 | 3 |
| ProducPhoto | 2 | 1 | 2 | 5 | 1 |
| Address | 5 | 5 | 5 | 15 | 4 |
| Employee | 7 | 6 | 3 | 16 | 5 |
| Product | 8 | 7 | 4 | 19 | 6 |

*RDI is the risk of data inconsistency, TC is table complexity, TZ is table size.

The following Fig 3 presents a plot of the ranking results on a matrix. The two axes correspond to the two ranking dimensions: normalization cost and quality impact. Looking at the figure, we can observe that tables that fall above the diagonal are the most promising tables to normalize since the impact rank higher than the cost. For example, table `WorkOrder` is potentially cheapest to normalize and have a likely high negative impact on software quality. Tables close to or on the diagonal tend to have balanced cost/impact ranking characteristics. For example, `Product` and `Employee` are the most expensive to normalize, but they also have high effect on quality characteristics. On the other hand, tables below the diagonal tend to be expensive to normalize and have little impact on the quality. These debts are likely to have low interest (i.e. low impact on quality) and high principal. For example, table `ProductProductPhoto` is likely to be more expensive to normalize but show only little negative impact. Normalizing this table can be deferred.

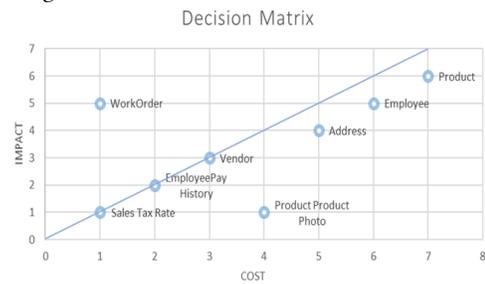

**Figure 3:** Normalization debt decision matrix

It is also possible to choose among tables to normalize based on their impact on a single quality dimension. For example, if the stakeholders are most concerned with risk of data inconsistency, the following Fig 4 illustrates a different decision matrix. It shows tables `Product` and `Employee` to be above the diagonal as they have the highest risk of data inconsistency but also the most costly to normalize. Similarly, Fig 5 and Fig 6 will demonstrate different decision matrixes based on table complexity and size respectively.

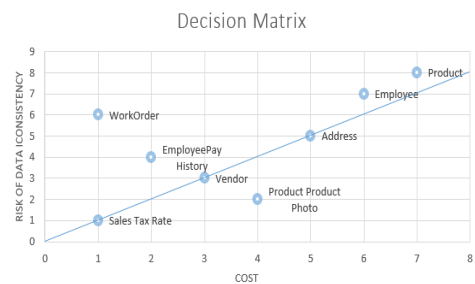

**Figure 4:** Decision Matrix based on cost and risk of data inconsistency

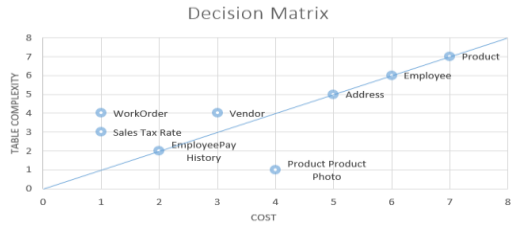

**Figure 5:** Decision Matrix based on cost and table complexity

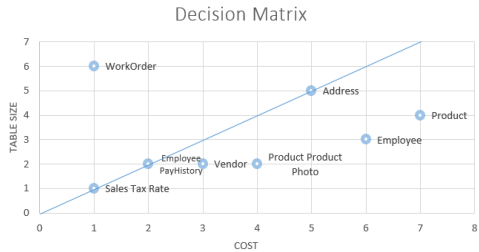

**Figure 6:** Decision Matrix based on cost and table size

The tradeoffs between the cost of normalization and the quality impact is based on the current information that is related to the size, complexity and risk of data inconsistency for a given time. However, as the database expect to grow by time and so the size, complexity and data inconsistency risk, the analyst can still use our technique to analyze for the decision of normalizing using the current information. If the analyst would be able to project that growth with some confidence, information related to the anticipated growth and the current state can be used as basis for tradeoffs against the decision of normalizing or not.

## 6 Evaluation

We have claimed that the fourth normal formal is the ideal/target normal form for more "debt-friendly" design. However, our case study has shown that designers should not think of the normal form as the only criteria justifying the normalization exercise and getting a "debt-friendly" design. In particular, the case has shown that issues that relate to cost and debts are to consider along the risk of data inconsistency, table size and complexity in the normalization exercise. Though the theory of database normalization could advocate restructuring into the fourth normal for all tables, our method has shown that designers can reduce efforts of unwanted/unnecessary normalization, which conventional approaches to normalization can still calls for. In particular, our method has provide better insight on justifying whether it is beneficial to go for 4$^{th}$ normal form based on the likely debts cost and quality ramifications that can be incurred (or cleared) from the exercise. Table 7 depicts the difference in normalization costs for tables under study using the conventional approach and our debt-aware approach. Results show that the debt-aware provide significant reduction in normalization efforts for four scenarios.

**Table 7:** Difference in cost between the conventional approach and the debt-aware approach of normalization

| Approach | Number of tables to normalize | Total Cost in $ |
|---|---|---|
| Conventional approach | 8 | 769.08 |
| Debt-aware approach | | |
| Option 1(Fig 3) | 1 | 51.36 |
| Option 2(Fig 4) | 4 | 532.86 |
| Option 3(Fig 5) | 3 | 166.92 |
| Option 4(Fig 6) | 1 | 51.36 |

Table 7 shows that the conventional approach, that follows aggressive normalization and calls for restructuring into the ideal fourth normal form, is more expensive and ad hoc in reaching this decision, as when compared to the debt-aware approach. The total cost is calculated using the method explained in section 5.1. In contrary, looking at the debt-aware approach; the approach can provide more options based on the scenario and developers preferences. Extrapolating these options from Decision Matrices of figures 3, 4, 5 and 6, where tables above the diagonals are recommended for normalization, we can see that Option 1 (Fig 3) suggests only one table, `WorkOrder`, based on the aggregated impact for dimensions related to the risk of data inconsistency, table complexity and size along the likely cost of normalization. Option 2, 3 and 4 examine the need to normalize based on one criteria of interest against the likely cost. For option 2, 3 and 4, the criteria are the risk of data inconsistency, complexity and size respectively. Interestingly, the four options recommends normalizing table `WorkOrder` as the baseline. Based on the priority and the available resources, the designer can include more tables along the recommended options. As observed, the decision matrices have provided an effective yet simple tool to visualize the tradeoffs between the cost and the likely quality impact of the candidate tables for normalization. Though option 2 of the debt-aware approach seems the most expensive recommendation calling for four tables including `WorkOrder`, it is still more cost-effective than aggressive normalization. In summary, the debt-aware approach had provided more systematic and informed cost-effective decision than the conventional approach.

### 6.1 Scalability and effort

Though the case study is hypothetical, the number of tables, the amount of data stored in each table and the relationship that exist between tables make the case a strong candidate for discussing scalability of the technique as it approximates real environments . In particular, the application of the method has benefited from manual and automated set up. For the manual, since our method gives a rough estimate of normalization cost, it requires experts input to quantify the cost of normalization and candidate tasks for refactoring. Though the conclusions on these dimensions could vary from one expert to another, designers are informed by experience and possibly good design patterns, which are suitable for the said domain. However, designers don't have quantitative means to decide on the likely benefits and costs of this exercise, which our approach provides.

Experts' judgment is also required when using the framework to identify normalization debt. In the case study, we assumed that the data is clean with no outliers due to accidental and erroneous updates. Henceforth, we have the sett of the confidence level to be 1. However, practically, accidental updates may occur in the data. This can be addressed by decreasing the confidence value of the extracted rules to less than 1. Once the extracted rules reports a confidence level less than 1, the tables can be manually inspected and reviewed by domain experts to detect errors and ensure validity of the functional dependency.

Despite the additional effort to set up the analysis and to quantify the likely impact and cost of normalization debt; the overhead for setting the quantitative analysis is justified as it likely to eliminate unwanted effort of normalizing tables that have minimum impact on the quality of system as shown in the previous section. The measurement of data inconsistency risk, complexity and size can be automated providing efficient alternative that support human judgement. This automation will not only save effort, it will facilitate wider range of what-if analysis.

## 7 RELATED WORK

Ward Cunningham was the first to introduce this metaphor in 1992 on the code level, as a trade of between short term business goals (e.g. shipping the application early) and long term goals of applying the best coding practices [9]. In [33], the authors compared human identification of technical debt to automatic identification using specific tools to analyze the code. They found that developers reported on different types of debts than the tools had elicited. Besides technical debt identification, debt management captured the attention for many researchers [16]. Our prioritization technique was inspired by the work done in [35], where the authors used cost/ benefit analysis to prioritize classes to refactor in the source code. They measured the effort cost of refactoring based on Marinescu's metrics of God classes (i.e. known as "classes that knows too much and do too much"). They argued that higher values of these metrics imply higher cost for refactoring. However, this argument was based on assumptions and is pending to validation as acknowledged by the authors. The authors measured the impact of the God classes on the systems maintainability and correctness. Finally, they prioritized classes to refactor based on the cost of refactoring and the quality impact of these classes to manage the design debts. Our work is different as it is driven by Normalization rules, which are fundamentally different in their requirements and treatment. Tables in databases have specific requirements and the likely impact of normalization can be observed on different quality metrics that are database specific. Therefore, the cost and debts needs to be discussed in the context of databases. Estimating technical debt principal was studied by many researchers [16]. The authors in [3] defined the principal as the time needed to fix debt prone bugs, and they used KNN-regression model to predict that time from historical data. Curtis et al. [11] presented a formula to calculate the principal of code and architectural level debt. Their calculation is based on the percentage of code and architectural violations to be fixed; time needed to fix each violation and the cost of labor. Despite the vast contribution on technical debt types, database design debt received very little attention. Weber et al. [30] discussed a specific type of schema design debt in missing foreign keys in relational databases. To illustrate the concept, the authors examined this type of debt in an electronic medical record system called OSCAR. The authors proposed an iterative process to manage foreign keys debt in different deployment sites.

Functional dependency detection has been discussed extensively [6], [22]. However, Mannila et al. [22] proved that discovering the complete set of dependencies within a single table is exponential in the number of attributes. Our framework for identifying functional dependencies avoids the exponential search space by testing only candidate dependencies between limited set of attributes rather than the whole table. Those candidate dependencies may violate a certain normal form criteria. If a violation is found, the process is terminated and the table is determined as a debt table.

## 8 CONCLUSION AND FUTURE WORK

We have explored a new concept of technical debt in database design that relates to database normalization. Normalizing a database is acknowledged to be an essential process to improve data quality, maintenance and flexibility. Conventional approaches for normalization is driven by the acclaimed structural and behavioral benefits that higher normal forms are likely to provide. The problem relies on the fact that developers tend to overlook this process for several reasons, such as, saving effort cost, lack of expertise or meeting the deadline. This can imply a debt that need to be managed carefully to avoid negative consequences in the future. Conversely, designers tend to embark on this exercise without clear justification for the effort and debt avoidance. We reported on novel method for identifying and managing the normalization debt in database design. We utilized association rule mining technique on the data stored in a table to discover dependencies between attributes and determine the normal form of the table and likely debt items. A table below the $4^{th}$ normal form is viewed as a potential debt item. Though we considered tables below the $4^{th}$ normal form as debt item, in practice most databases lag behind this level [15] making it impractical to use it as the only criteria to drive the normalization exercise. To overcome this problem, we propose a method to prioritize debt items and their candidacy for normalization based on the cost of normalization (i.e. debt principal), and the likely impact of the debt tables on the quality (i.e. debt interest), including risk of data inconsistency, complexity and size of each debt table. The method was applied to AdventureWorks database from Microsoft that approximates real world environment and consists of 64 tables, exhibiting complex dependencies and populated with large amount of data. The results show that rethinking conventional database normalization from the debt angle, by not only relying on the normal form to re-design the database, can provide more systematic guidance and informed decisions to improve the quality while reducing the cost of the normalization. For future work, we intend to incorporate more quality measures in prioritizing tables need to be normalized, such as, portability and availability and we hope to transit the analysis to many objectives optimization for debts and qualities.